\documentclass[twocolumn]{aastex631}

\newcommand{\angstrom}{\mbox{\normalfont\AA}}
\usepackage{soul}

\usepackage{CJK}
\usepackage{amsmath}

\received{April 10, 2023}
\revised{July 5, 2023}
\accepted{July 28, 2023}

\begin{document}

\title{Detection of the Low-Stellar Mass Host Galaxy of a $z\sim6.25$ Quasar with JWST}

\correspondingauthor{Meredith Stone}
\email{meredithstone@arizona.edu}

\author[0000-0002-9720-3255]{Meredith A. Stone}
\affiliation{Steward Observatory, University of Arizona, 933 North Cherry Avenue, Tucson, AZ 85721, USA}

\author[0000-0002-6221-1829]{Jianwei Lyu (\begin{CJK}{UTF8}{gbsn}吕建伟\end{CJK})}
\affiliation{Steward Observatory, University of Arizona,
933 North Cherry Avenue, Tucson, AZ 85721, USA}

\author[0000-0003-2303-6519]{George H. Rieke}
\affiliation{Steward Observatory, University of Arizona,
933 North Cherry Avenue, Tucson, AZ 85721, USA}

\author[0000-0002-8909-8782]{Stacey Alberts}
\affiliation{Steward Observatory, University of Arizona,
933 North Cherry Avenue, Tucson, AZ 85721, USA}

\begin{abstract}

We characterize the stellar mass of J2239+0207, a $z\sim6.25$ sub-Eddington quasar (M$_{1450} = -24.6$), using dedicated JWST/NIRCam medium-band observations of a nearby PSF star to remove the central point source and reveal the underlying galaxy emission. We detect the host galaxy in two bands longward of the Balmer break, obtaining a stellar mass of $\sim10^{10}$ M$_{\odot}$, more than an order of magnitude less than this quasar's existing measured [C II] dynamical mass. We additionally calculate the mass of J2239+0207's central supermassive black hole using JWST/NIRSpec IFU observations, and determine that the black hole is $\sim15$ times more massive than predicted by the local $M_{\mathrm{BH}}$-$M_{*}$ relation, similar to many high-redshift quasars with dynamical masses determined via millimeter-wave line widths. We carefully consider potential selection effects at play, and find that even when $z\sim6$ quasars are compared to a local sample with similarly determined dynamical masses, many of the high-redshift quasars appear to possess overmassive black holes. We conclude $z\sim6$ quasars are likely to have a larger spread about the $M_{\mathrm{BH}}$-$M_{*}$ relation than observed in the local Universe.

\end{abstract}

\section{Introduction} \label{sec:intro}

Active galactic nuclei (AGN) are energetic enough to greatly impact the star formation, morphologies, and other properties of their host galaxies, but the nature of these effects---especially at high redshift---is still poorly understood. Understanding the interplay between galaxies and their central supermassive black holes (SMBHs), including their coeval growth and triggering of the AGN phase, is critical to building a complete picture of galaxy formation and evolution.

Studying powerful AGN---particularly quasars---{\em and} their host galaxies is difficult. Black hole masses are easiest to measure up to high redshift in Type 1, face-on AGN via single-epoch spectra, where the Doppler-broadened emission lines from gas orbiting the black hole are visible in the spectrum. However, in these systems the bright quasar tends to outshine the emission from its host, making the characterization of the host galaxy challenging.

At high redshift, the properties of quasar host galaxies can be explored via the rest-frame far-infrared and submillimeter regimes, where the contribution from the AGN is low and the emission is dominated by cold gas and dust tracing regions of star formation \cite[e.g.,][]{Lyu2016}. The width of the [C~II] $158 \mu$m fine-structure cooling line in particular, accessible at high redshift with ALMA and other ground-based submillimeter telescopes, can be used to probe the dynamical mass of quasar hosts \citep[e.g.][]{Wang2013, Venemans2016, Decarli2018, Izumi2019} in systems where the stellar emission is difficult to distinguish from the quasar \citep{Shao2017, Pensabene2020, Neeleman2021, Shao2022}. However, this method is subject to a number of assumptions and uncertainties regarding the nature of the galaxy and its orientation on the sky. To improve on this approach, submillimeter observations can be used to create spatially-resolved line maps and model the dynamics of the host galaxies. Dynamical masses determined in these ways are often compared with local relations based on more directly determined stellar masses when comparing high-redshift quasar samples to the local Universe via e.g. the $M_{\mathrm{BH}}/M_*$ relation \citep{Izumi2019, Pensabene2020, Yue2021}.

Another, more direct method to detect the underlying stellar emission from quasar host galaxies and estimate their stellar masses is to subtract a theoretical or empirical point-spread function (PSF) from quasar images. However, these efforts have mainly been limited to $z \lesssim 2-3$ \citep[e.g.][]{Gehren1984, Hutchings2002, Dunlop2003, Jahnke2004, Kim2017, Yue2018, Li2021}, using ground-based optical telescopes or the Hubble Space Telescope to probe the rest-frame optical to near-infrared emission of the quasar where the host galaxy contribution is the strongest. These studies have revealed that SMBH masses ($M_{\mathrm{BH}}$) and the masses of their host galaxies ($M_{*}$, or the central bulge mass $M_{\mathrm{bulge}}$) are correlated in the low- and intermediate-redshift Universe \citep{Kormendy2013, Reines2015}. The question follows: does this correlation persist to high redshift? A steeper relation at high-z---i.e. overmassive black holes relative to their host galaxies---implies that black holes build up their masses before their host galaxies do, whereas greater scatter at high-z implies that the coeval growth of supermassive black holes and galaxies has not yet settled to its local behavior.

Most of the small number of distant ($z \gtrsim 3$) quasars whose host galaxy emission has been detected or constrained \citep[e.g.][]{Peng2006, McLeod2009, Mechtley2012, Targett2012, Schramm2019, Marshall2020, Marshall2023} lie above the established $M_{\mathrm{BH}}$-$M_{*}$ relation, with overmassive black holes given their stellar mass; on the other hand, studies of high-$z$ quasars with dynamical masses measured from [C~II] have implied $M_{\mathrm{BH}}$-$M_{*}$ relations both above and consistent with the local Universe \citep{Walter2004, Wang2013, Venemans2016, Decarli2018, Izumi2019, Farina2022}. However, this observed trend may be due in part to selection effects: at such high redshifts, studies are biased towards common, lower-mass galaxies with overmassive black holes \citep{Lauer2007, Zhang2023}. Additionally, at these redshifts, optical telescopes probe rest-frame ultraviolet emission of galaxies, rather than the rest-frame optical that directly traces the stellar mass. It is therefore important to study underluminous quasars accreting below the Eddington limit---and in particular their rest-frame optical emission--- to mitigate this bias and explore the potential scatter of the high-redshift $M_{\mathrm{BH}}$-$M_{*}$ relation.

JWST's NIRCam instrument \citep{Rieke2005, Rieke2023} will greatly expand the number of high-redshift quasars with host galaxy detections by probing stellar emission out to $z \sim 9$ at high sensitivity and resolution. High-$z$ quasar samples being built with NIRCam will help to fill out the high-$z$ $M_{\mathrm{BH}}$-$M_{*}$ relation to better understand the details of galaxy formation and evolution in the first $\sim1$ Gyr after the Big Bang \citep[see e.g.][for other early results]{Ding2022}. To demonstrate this possibility, we designed a JWST GTO program (ID 1205; \citealt{Rieke2017}) to obtain NIRCam imaging and/or NIRSpec IFU observations of six $z\sim5-6.5$ quasars spanning a range of host galaxy star formation rate, AGN luminosity, and obscuration. 

In this paper, we report our first results for the $z\sim6.25$ quasar HSC J2239+0207 with multi-band NIRCam imaging and NIRSpec prism IFU observations. Discovered by the Subaru High-$z$ Exploration of Low-Luminosity Quasars (SHELLQ) project \citep{Izumi2019}, J2239+0207 is a moderately-luminous quasar ($M_{\rm 1450, optical}=-24.7$ or $L_{\rm bol}\sim2.5\times10^{12}~L_\odot$) that is accreting at $\sim$30\% of the Eddington limit, though its precise Eddington ratio is uncertain due to the spread in measured BH masses: $M_{\mathrm{BH}}=6 \times 10^8$ and $1 \times 10^9 \; M_{\odot}$ from C~IV $\lambda1549$ and Mg~II $\lambda2798$ respectively \citep{Onoue2019}. The quasar's redshift situates the stellar $4000 \, \angstrom$ break at approximately 3 $\mu$m, amidst the long-wavelength NIRCam filters. Due to its relatively low Eddington ratio and high dynamical mass \citep[$\log \,$M$_{\mathrm{dyn}}$ (M$_\odot)$$ = 11.5$,][]{Izumi2019}, the host galaxy of J2239+0207 should be easily detectable in the rest-frame optical. J2239+0207 is therefore an ideal case for host galaxy detection and characterization of a quasar at the end of the Epoch of Reionization with JWST.

This paper is organized as follows. Section~\ref{sec:data} summarizes the JWST observations of J2239+0207 and the relevant data reduction processes. We outline the procedures to extract the host galaxy signal from the NIRCam images and to determine the black hole mass from the NIRSpec spectrum, and present our measurements in Section~\ref{sec:results}.  After characterizing its host galaxy mass, we determine the location of J2239+0207 relative to the local $M_{\mathrm{BH}}$-$M_{*}$ relation and, considering various selection biases, discuss its implications on the cosmic coevolution of supermassive black holes and their host galaxies in Section \ref{sec:discussion}. A final summary is given in Section~\ref{sec:summary}. Throughout this work, we assume $H_0$ = 69.6, $\Omega_M$ = 0.286, and $\Omega_{\Lambda}$ = 0.714.

\section{Observations and Data Reduction} \label{sec:data}

\subsection{NIRCam}

HSC J2239+0207 was observed in three NIRCam medium-band filters, with total exposure times of 2227.2 seconds in F210M and 1113.6 seconds in F360M and F480M. We used Module B SUB400P, (FOV 12.5$\arcsec\times12.5\arcsec$, SW; $25\arcsec\times25\arcsec$, LW), to image the quasar field using the BRIGHT2 readout mode. To improve the PSF sampling and mitigate cosmic rays and detector artifacts, we adopted 4$\times$4 primary and sub-pixel dithering patterns with dither types INTRAMODULEBOX and STANDARD, respectively. To obtain a reference PSF, the nearby bright star 2MASS 2M22390990+0207329 was also imaged in the same filters and observing setup with the RAPID readout mode and shorter exposure times to avoid saturation (424 seconds in F210M, 265 seconds in F360M, and 159 seconds in F480M). We selected these medium-band filters to minimize changes in the point spread function (PSF) due to the different shapes of the spectral energy distributions of the quasar and star across the photometric bandpass: differences in the color of the quasar and star have negligible effect (see Section \ref{subsec:psf}). 

\begin{figure*}
    \centering
    \includegraphics[width=18cm]{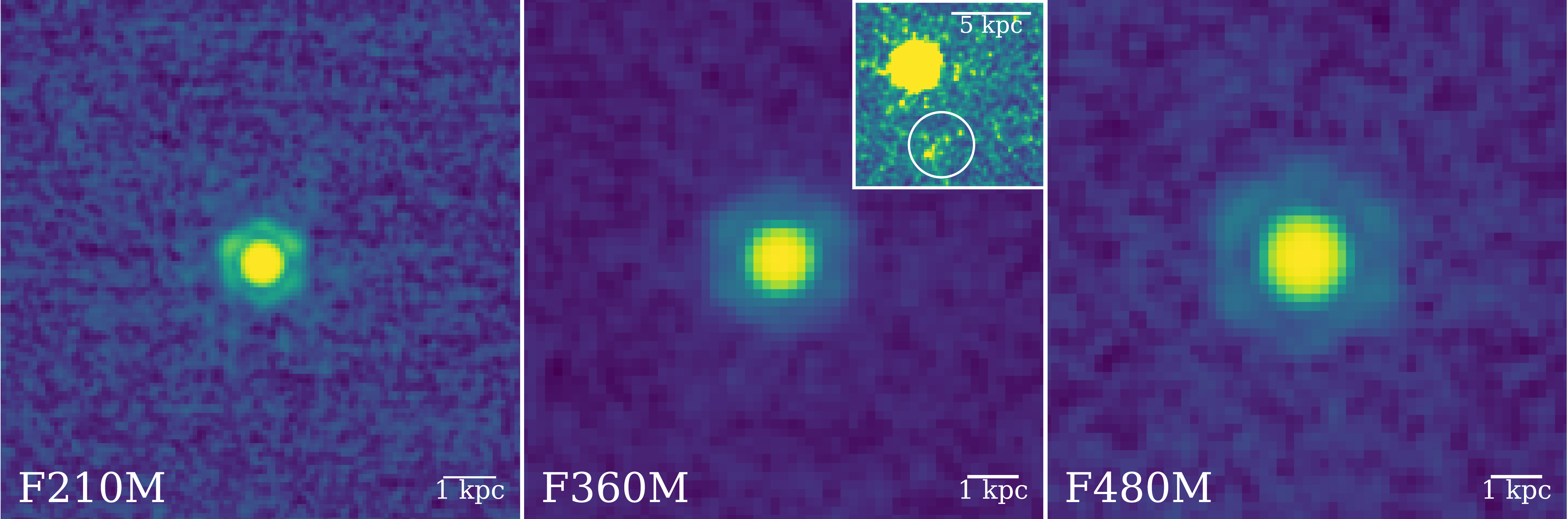}
    \caption{NIRCam images of J2239+0207 in our three science bands: F210M {\em (left)}, F360M {\em(center)}, and F480M {\em(right)}. All three images have a field of view of $\sim1.8$\arcsec. The F360M image, with a different stretch and greater field of view, is shown inset into the center panel to show the emission-line companion (circled) visible in both our NIRSpec spectrum and NIRCam image at a distance of $\sim1\arcsec$ from the quasar.}
    \label{fig:quasar}
\end{figure*}

We processed this data using the JWST pipeline version 1.8.4, following roughly the standard procedures as recommended by the STScI JWebbinars\footnote{\url{https://github.com/spacetelescope/jwebbinar_prep/tree/main/imaging_mode}}. The pipeline parameter reference file is jwst\_1015.pmap, as registered in the JWST Calibration Reference Data System (CRDS). During Stage 2, we added a custom step to measure and subtract the image striping caused by the 1/f noise from the detector readout in each frame. The final mosaic images were produced by aligning and stacking all the frames obtained in Stage 2 using the Stage 3 pipeline. To improve the accuracy of PSF subtraction, these images were re-sampled with smaller pixel scales at 0.0147$\arcsec$/pixel in F210M, and 0.0300$\arcsec$/pixel in F360M and F480M with the drizzling algorithm in the Resample step. Finally, we conducted a global background subtraction using the Photutils Background2D function. The background-subtracted quasar images in the three bands are shown in Figure \ref{fig:quasar}: all three bands are dominated by the central PSF.

\subsection{NIRSpec}

To aid the analysis and interpretation of the host galaxy signals, we also obtained NIRSpec IFU data of J2239+0207 with the PRISM/CLEAR disperser-filter combination. This IFU data has a $3\arcsec\times3\arcsec$ field of view with the size of each IFU element at $0.1\arcsec\times0.1\arcsec$, and wavelength coverage from 0.6 $\mu$m to 5.3 $\mu$m with nominal spectral resolution $\sim$100 but increasing significantly toward longer wavelengths. To mitigate cosmic ray and detector artifacts, we adopted the SPARE-CYCLING dither type in LARGE size at Points 1, 2, 3, 4. With the NRSIRS2 readout, the total integration time was 8870 seconds. 

We reduced the NIRSpec IFU data using the same JWST pipeline version and CRDS file as the NIRCam data. Initially we followed the standard pipeline,\footnote{\url{https://github.com/spacetelescope/jwebbinar_prep/tree/main/ifu_session}} but found it has several issues: for example, a few IFU elements at the quasar center had been mistakenly identified as outliers by the default pipeline and automatically masked out. We therefore performed a visual inspection of the data cube before and after the outlier detection to differentiate the true outliers from the mistakenly flagged bright quasar pixels. We produced a final data cube for science analysis by skipping the automatic outlier detection step, and instead manually masking out the IFU elements we believe to be true outliers.

\begin{figure*}
    \centering
    \includegraphics[width=0.7\hsize]{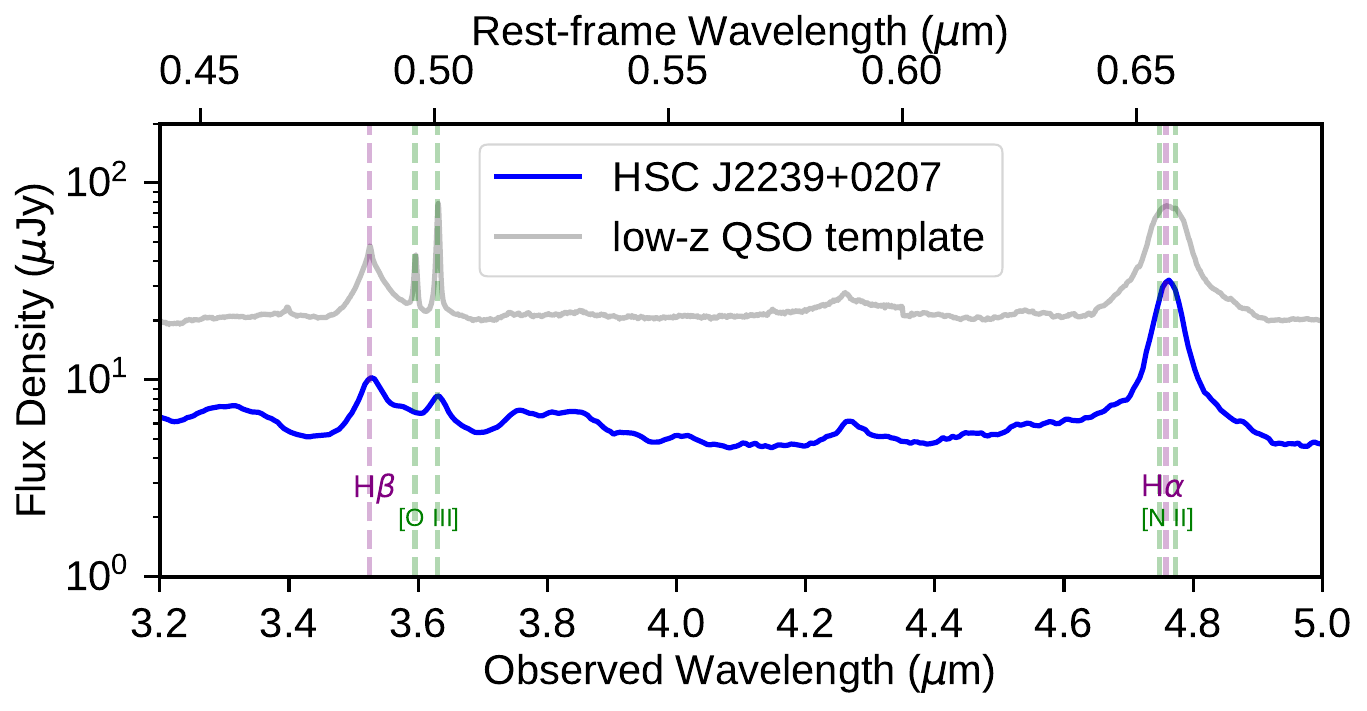}
    \caption {Observed NIRSpec spectrum of the HSC J2239+0207 nucleus with some prominent emission lines identified. As a comparison, we also plot
    an average spectral template derived from the low-$z$ quasars in the DR7 edition of the SDSS Quasar Catalog \citep{Schneider2010}.
    }
    \label{fig:spectrum}
\end{figure*}

In this work, we use the IFU data to measure the quasar black hole mass and determine if any host galaxy signal detected via PSF subtraction of the NIRCam data is strongly contaminated by gas emission lines. We therefore extract a quasar spectrum using a circular aperture at a radius of 0.35$\arcsec$, as shown in Figure~\ref{fig:spectrum}. Further analysis of the IFU data is reserved for a future paper.

\section{Analysis and Results} \label{sec:results}

\subsection{Retrieving Host Galaxy Signals via PSF Subtraction} \label{subsec:psf}

\begin{figure*}
    \centering
    \includegraphics[width=16cm]{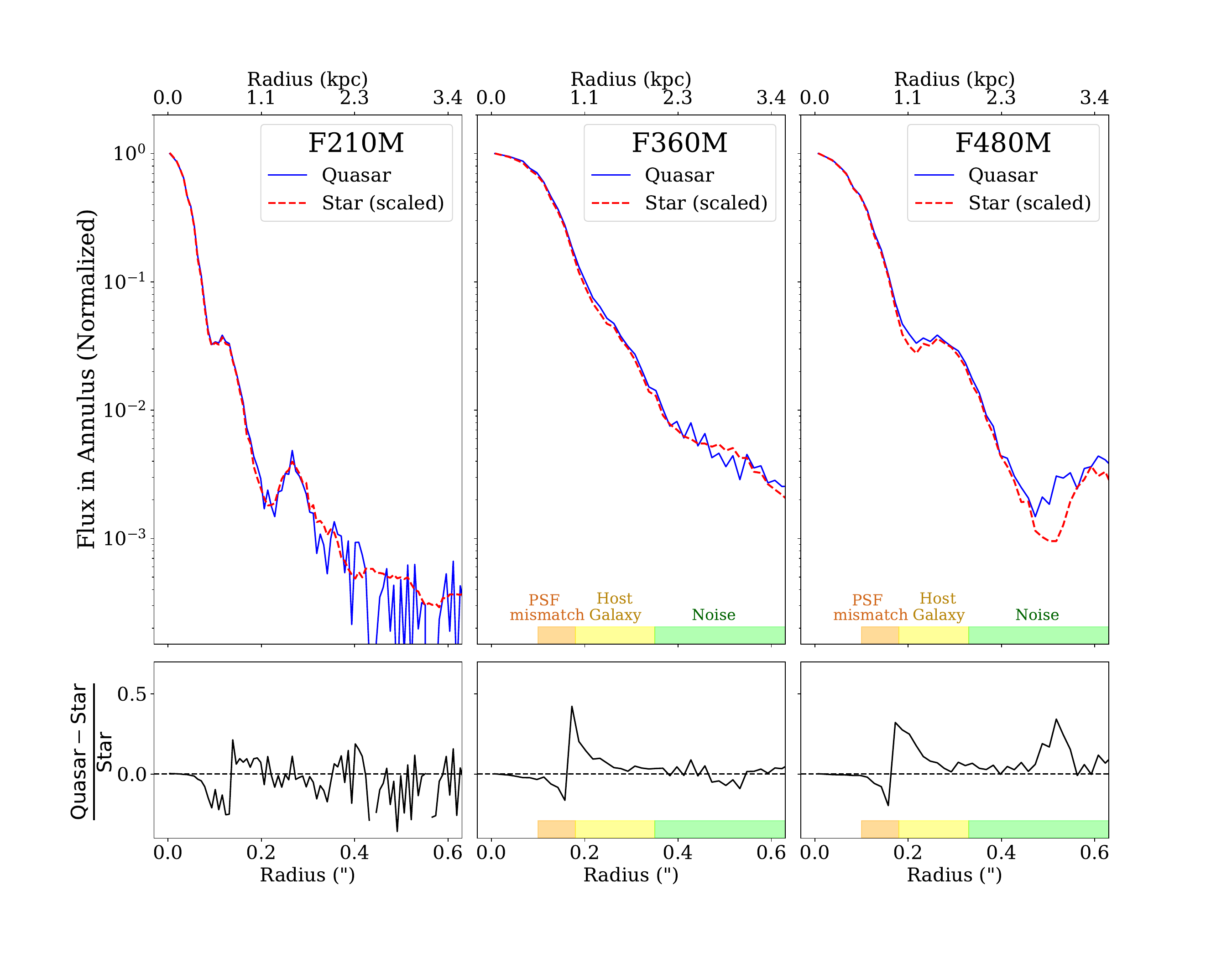}
    \caption{Comparison of the observed radial profiles of J2239+0207 and the PSF reference star in our three bands. {\em Top:} The radial profiles of the PSF star (red dashed line) and quasar (solid blue line): the star image has been scaled to match the quasar flux in the central pixels. The pixel sizes are 0\farcs015 for F210M and 0\farcs030 for the other two bands. The quasar PSF does not appear broader than the stellar PSF in any band; the risk of a spurious galaxy detection due to PSF mismatch is therefore low. {\em Bottom:} The normalized difference in flux between the star and quasar profiles. Slight mismatches between the shape of the quasar and stellar PSFs near the core produce the dip and spike observed in F360M and F480M (orange highlighted region). The host galaxy is visible via the excess flux at a radius of $\sim 1-1.5$ kpc/$\sim0.2 \arcsec$ in both F360M and F480M, but not in F210M (yellow highlighted region). Finally, at radii greater than $\sim 0.4\arcsec$, noise dominates and any excess observed is likely not meaningful (green shaded region).}
    \label{fig:profiles}
\end{figure*}

As there are not a large number of bright stars in the quasar field suitable for building an empirical PSF, we obtained a dedicated observation of a PSF star. The NIRCam images of the star were taken immediately after observing the quasar to minimize temporal variation of the PSF between the observations. We take the stellar image as the PSF and scale and subtract it from the quasar directly, rather than building an empirical PSF from stars in the image, or fitting a model that relies on assuming certain properties of the underlying host galaxy. \cite{Wolff2023} evaluated a number of state-of-the-art PSF subtraction techniques in a search for the debris disk around $\epsilon$ Eridani, and find this technique to be optimal. The PSF star and quasar are also placed at the same location on the detector, which minimizes PSF differences between the star and quasar due to distortions across the detector. Any differences in sub-pixel alignment between the star and quasar PSFs are mitigated by the dithering process.

Because the quasar and the PSF star have different spectral shapes within the bands (the relatively flat optical spectrum of the quasar compared to the declining Rayleigh-Jeans tail of the star) it is possible that the shapes of their PSFs will differ. This effect will tend to introduce a spurious galaxy detection, as the quasar's redder spectrum will produce a broader PSF than the star. The choice of NIRCam medium bands for this analysis helps to mitigate this effect, but we also investigate any possible differences by comparing NIRCam PSFs generated with WebbPSF \citep{Perrin2014} using as input a Rayleigh-Jeans spectrum and our NIRSpec spectrum of J2239+0207. The resulting WebbPSFs display functionally identical radial profiles in F360M and F480M at all radii, and are very similar in F210M: the median normalized difference ($\frac{\mathrm{Quasar}-\mathrm{Star}}{\mathrm{Star}}$) between the quasar and star profiles at all radii is $\sim3\%$ in F210M and $\sim0.5\%$ in F360M and F480M. The difference in flux obtained within a 0.4\arcsec-radius aperture between the simulated star and quasar is $\sim 0.3\%$ in F360M and F480M and $\sim 1.5\%$ in F210M. We can therefore confirm that the differences in the PSF of the quasar and star have a negligible effect on this analysis, particularly away from the PSF core.  

\begin{figure*}
    \centering
    \includegraphics[width=18cm]{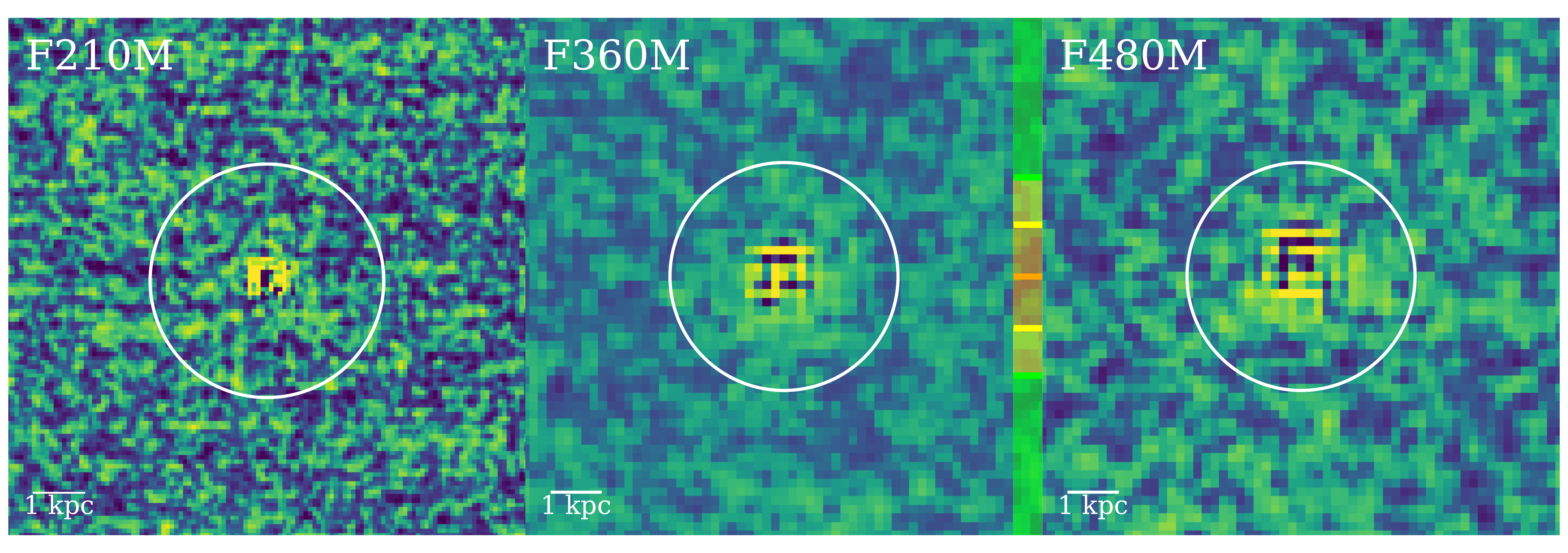}
    \caption{The quasar image in F210M (left), F360M (center) and F480M (right) after subtraction of the stellar PSF, normalized by the total (Poissonian + pixel) error budget. The $0.4\arcsec$-radius aperture used to extract the host galaxy flux is shown in white. All three images are shown with the same scaling and stretch. F210M does not display significant emission beyond the central PSF artifacts, but extra extended emission is visible in F360M and F480M. Orange, yellow, and green shaded regions, as in Figure \ref{fig:profiles}, mark the radii at which the emission is dominated by PSF artifacts, host galaxy emission, and noise, respectively. All images have a field of view of $\sim 1.8$\arcsec.
    }
    \label{fig:galaxy_companion}
\end{figure*}

To subtract off the quasar point-spread function from the images, we first align the background-subtracted quasar and PSF star images in each band. We normalize the stellar PSF image in each band, scaling it to match the flux of the star to the flux of the quasar in the central pixels, where the quasar flux is dominated most heavily by the point source rather than any underlying galaxy emission. We subtract the scaled star image from the quasar image and examine the residuals, verifying that this simple method of normalizing the PSF does not leave obvious over- or under-subtraction artifacts in the resulting PSF-subtracted images (see Figure \ref{fig:galaxy_companion}). 

The radial profiles of the quasar and the star (the latter scaled to match the quasar in the central pixels) in all three bands are shown in Figure \ref{fig:profiles}. The quasar and stellar profiles are very similar within $\sim2$ kpc of the center of the image where we expect to observe galaxy emission, with only slight evidence of a broader quasar PSF relative to that of the star very near the PSF core, consistent with the WebbPSF predictions. However, the quasar F360M and F480M profiles display low-level excess emission ($\sim 10$\% of the quasar signal at the relevant radii) at a distance of $\sim1$ kpc or 0.2\arcsec (bottom panels); no such strong excess is seen in F210M. For comparison, the maximum deviation at any radius between the simulated WebbPSF quasar and stellar radial profiles was $\sim 7\%$ in F360M and F480M.

The PSF-subtracted quasar images are shown in Figure \ref{fig:galaxy_companion}. F210M, which at the quasar redshift probes shortward of the 4000\angstrom $\,$break, shows bright PSF subtraction artifacts at the center but exhibits no evidence of underlying galaxy emission, consistent with the match of the quasar and star radial profiles in the leftmost panel of Figure \ref{fig:profiles}. As hinted by the radial profiles, some faint signal is visible around the quasar in F360M and F480M, which both lie longward of the Balmer break. We measure galaxy fluxes within a circular 0.4$\arcsec$-radius ($\sim 2$ kpc) aperture of 0.34 $\mu$Jy in F360M and 0.44 $\mu$Jy in F480M (6.8 and 5.5$\sigma$ respectively), which correspond to respective host-to-quasar flux ratios of 4.6\% and 3.9\%. We can also put a secure $3\sigma$ upper limit on the F210M flux: all fluxes are listed in Table \ref{tab:measurements}.

These fluxes likely slightly underestimate the flux of the host galaxy, because the galaxy flux near the PSF core is subtracted along with the PSF. In Section \ref{subsec:hostgalaxy}, we use the outskirts of the galaxy where the PSF artifacts are not severe to model the galaxy as a Se\'rsic profile and determine the degree to which we are underestimating the host galaxy flux.

\begin{deluxetable}{ccccc}
\tablecaption{Measured quasar and galaxy properties for J2239+0207 \label{tab:measurements}}
\tablewidth{0pt}
\tablehead{
\colhead{Band} & \colhead{Quasar} & \colhead{Galaxy} & \colhead{Approximate} & \colhead{Inferred} \\[-0.25cm]
\colhead{} & \colhead{Flux} & \colhead{Flux} & \colhead{Galaxy Size\tablenotemark{a}} & \colhead{Mass} \\
\colhead{} & \colhead{($\mu$Jy)} & \colhead{($\mu$Jy)} & \colhead{\arcsec} & \colhead{$\log$(M$_{\odot}$)}
}
\decimalcolnumbers
\startdata
F210M & $6.81 \pm 0.04$  & $< 0.15$ & N/A & N/A  \\
F360M & $ 7.43 \pm 0.05 $  & $0.34 \pm 0.05$ & $\sim1$ kpc & $10.0^{+0.3}_{-0.5}$ \\
F480M & $ 11.32 \pm 0.09 $ & $0.44 \pm 0.09$ & $\sim1$ kpc & N/A \\
\enddata
\tablenotetext{a}{Effective radius of best-fit S\'ersic profile (Section \ref{subsec:hostgalaxy})}
\end{deluxetable}

A companion to the quasar at a distance of $\sim 1$\arcsec is visible in F360M (see Figure \ref{fig:quasar}, central panel inset). A preliminary spectrum extracted from the NIRSpec data
shows this companion only in emission lines. Its redshift relative to the quasar suggests that it is either an outflow, or a galaxy passing close by. More details of this companion will be presented in a future paper.

\subsection{Black hole mass from the NIRSpec spectrum} \label{subsec:bhmass}

 To measure the emission lines in the quasar spectrum, we adopt a continuum based on the classic Elvis general result \citep{Elvis1994}, slightly adjusted in slope to match the slope of the quasar's continuum. It provides a good fit, with no indication of abnormalities, and is a useful baseline against which to identify emission lines. Prominent lines are indicated in Figure~\ref{fig:spectrum}. 

\begin{figure}
    \centering
    \includegraphics[width=9cm]{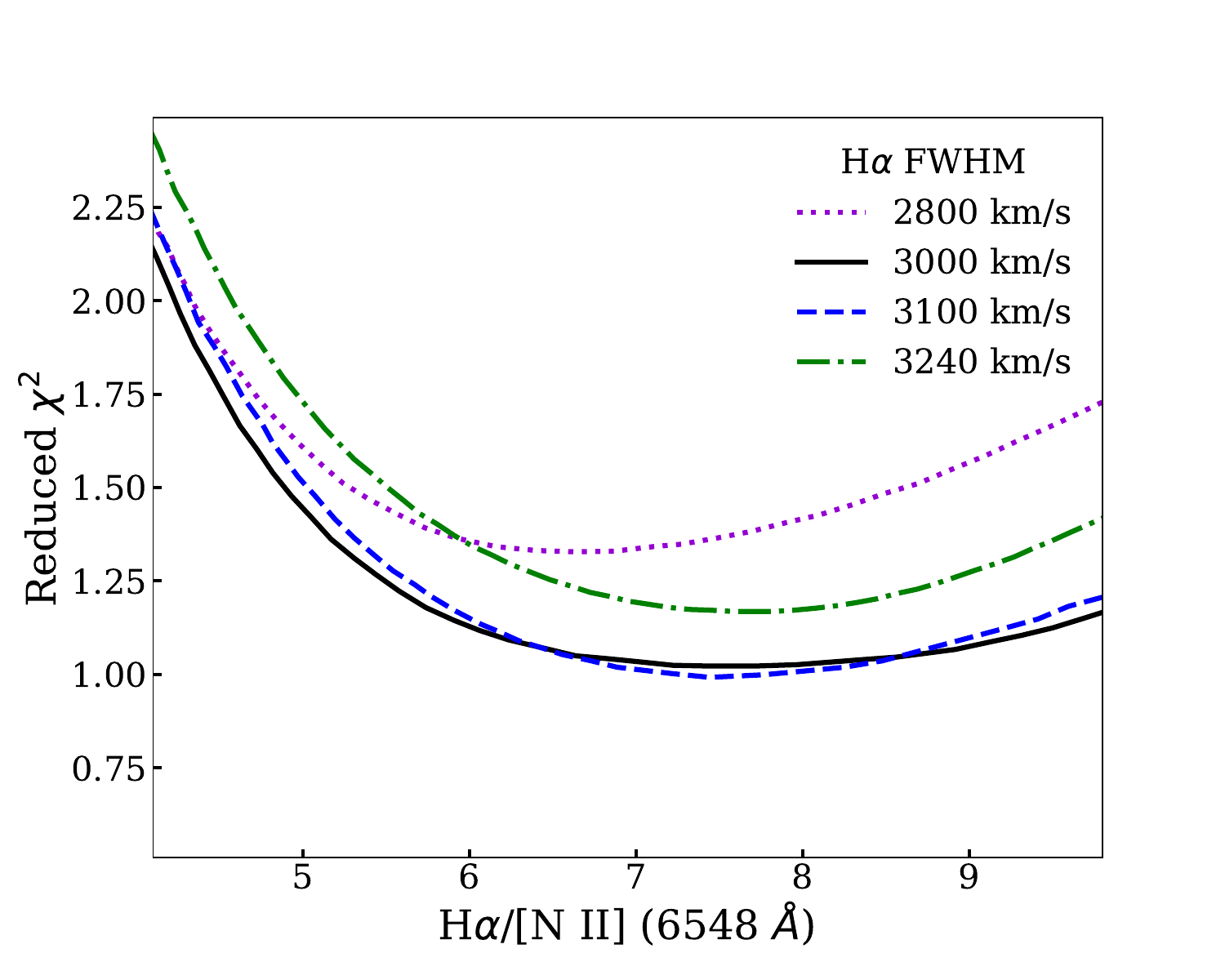}
    \caption{Results from modeling the H$\alpha$ - [NII] line blend. The figure shows the normalized $\chi^2$ as a function of the ratio of peak line strengths for H$\alpha$ to [NII]6548\AA. H$\alpha$ must be about seven times stronger than the shorter wavelength [NII] line and the preferred H$\alpha$ FWHM is about 3050 km s$^{-1}$.
    }
    \label{fig:linemodel}
\end{figure}

The H$\alpha$ line in our spectrum (Figure~\ref{fig:spectrum}) can provide an independent estimate of the black hole mass (the spectral resolution at H$\beta$ is too low to resolve the line well, and our efforts to model it were unsuccessful). At the observed wavelength of H$\alpha$, the prism spectrum has a resolution\footnote{https://jwst-docs.stsci.edu/jwst-near-infrared-spectrograph/nirspec-instrumentation/nirspec-dispersers-and-filters} of $\sim$ 1200 km s$^{-1}$. The FWHM of the blended H$\alpha$ line is $\sim$ 3000 km s$^{-1}$, so the spectral resolution is adequate to measure its width. To evaluate the contribution of the [N~II] lines to the width of H$\alpha$, we build a model of the three lines (H$\alpha$ and both [N~II] lines), with the width of the [N~II] lines set to the spectral resolution and the ratio of their fluxes to the theoretical value of 3. By varying the width of the H$\alpha$ line and its strength relative to the [N~II] lines, we can compare the resulting line profile to our observed spectrum and calculate a $\chi^2$ value. 
The results are shown in Figure~\ref{fig:linemodel}. The preferred width is $\sim 3050 \pm 250$ km s$^{-1}$. To check this result, we also measure the half width at half maximum blueward of the line center assuming $z = 6.2497$ \citep{Izumi2019}. Because the $\lambda$ 6543 line is at least six times weaker than H$\alpha$ (see Figure~\ref{fig:linemodel}), we expect the half width at half maximum not to be significantly affected by blending. The result is 1450 km s$^{-1}$, or a full width of 2900 km s$^{-1}$, in agreement within the expected errors. 

We adopt 3050 km s$^{-1}$. Removing the broadening due to the spectral resolution yields a final width of 2800 km s$^{-1}$. There may be a narrow component to the H$\alpha$ line---at our resolution there is no way to tell. The presence of a narrow component would make our width estimate too small: this width is therefore a lower limit. The single-epoch black hole relation is calibrated with H$\beta$, and there are subtle differences in the typical measured FWHMs of the two lines. We use the relation in \citet{Greene2005} (their equation 3) to convert to an equivalent H$\beta$ FWHM of $\gtrsim$ 3090 km s$^{-1}$. We can then use the relations in \citet{Vestergaard2006} (their equation 5) to convert the line widths to an estimate of the black hole mass, as:

\begin{multline}
    \log{M_{\mathrm{BH}}} = \log \Biggl \{ \! \biggl [ \frac{\mathrm{FWHM(H}\beta)}{1000 \; \mathrm{km \, s}^{-1}} \biggr ]^{2}   \biggl [ \frac{\lambda L_{5100 \mbox{\scriptsize\normalfont\AA}}}{10^{44} \; \mathrm{erg \, s}^{-1}} \biggr ]^{0.5} \Biggr \} \\ + (6.91 \pm 0.02)
\end{multline}
\noindent
We take the continuum luminosity at 5100 \angstrom $\,$ from the flux density in F360M, 7.34 $\mu$Jy, and correct it for the emission line contribution using the spectrum. The resulting mass is $\gtrsim (3.5 \pm 0.7) \times 10^8$ M$_\odot$. 

\citet{Onoue2019} estimated the mass of the black hole in  J2239+0207 using both the Mg~II and C~IV lines. They measured the FWHM of the lines to be, respectively, $4670^{+910}_{-700}$ and $4630^{+1040}_{-1260}$ km s$^{-1}$, corresponding to mass estimates of $11^{+3}_{-2}$ and $6.3^{+2.0}_{-2.5}$ $\times 10^8$ M$_\odot$. The latter value includes a correction for outflows indicated by the blueshift of the line \citep{coatman2017}; without this correction the value would be $8.9^{+2.8}_{-3.4} \times 10^8$ M$_\odot$. They also find a redshift of z = 6.2499. Taking the three mass estimates together (3.5, 6.3, and 11 $\times 10^8 \; M_\odot$), for a net value of  $\sim6 \times 10^8$ M$_\odot$, the Eddington ratio is $\sim 0.3^{+0.2}_{-0.15}$ (for L$_{bol} = 2.5 \times 10^{12}$ L$_{\odot}$)\footnote{Estimated from M$_{1450}$ \citep{Izumi2019}}. As shown by the differences among the three values, the uncertainty is dominated by systematic errors.

\section{Discussion} \label{sec:discussion}

\subsection{Host Galaxy} \label{subsec:hostgalaxy}

A SMBH of the mass discussed above would be expected to lie within a massive host galaxy. Assuming the local relation between black hole and galaxy bulge mass, the J2239+0207 black hole of $6 \times 10^8$ M$_{\odot}$ predicts a galaxy of $1.2 \times 10^{11}$ M$_\odot$ \citep{Kormendy2013}.

To compare with the values from our observations, we need to relate host galaxy mass to flux. Our low SNR detections of the host galaxy and small number of filters are inadequate to support detailed  modeling of its characteristics via e.g., spectral energy distribution fitting. However, we do test to see if our measurements are consistent with a plausible SED. The host galaxy has a strong far infrared excess \citep{Izumi2019}, which indicates a SFR of $\sim$ 280 M$_\odot$/yr assuming a Kroupa IMF \citep{Kennicutt2012}: the galaxy is in a starburst stage. We run a model\footnote{Our simulations utilize the PopStar models \citep{molla2009} for a Chabrier IMF, which
we combine to provide SEDs for different SFHs. We have modified the code to include emission lines, tied to the level of star formation. These models
do not include massive binary stars, which would extend the
duration of strong ionizing fluxes by a factor of $\sim 1.5 - 2$ \citep[e.g.][]{Stanway2016}, but would not change our results in a
significant way.} assuming the galaxy is halfway through a 20 Myr long burst that will account for a 10\% increase in its mass, which was built up by similar bursts distributed uniformally over the past 500 Myr. The measurements are consistent with the predicted SED provided the extinction is $A_V \ge 0.5$ mag. This model is far from unique; for example, if the high rate of star formation has a duration of 2 Myr rather than 10 Myr, then models with $A_V >$ 0.2 are consistent with the measurements.

As illustrated by these examples, model-fitting to our three measurements is highly degenerate. Two detections of moderate signal to noise and one upper limit are just too limited to constrain models further. We therefore  use observations of massive galaxies at similar redshift but without active nuclei to provide insights to the probable host galaxy properties. These galaxies should include the appropriate range of star formation histories and levels of extinction. In addition, many of them are measured  in {\em Spitzer}/IRAC Band 1, which coincides closely with our F360M observations. The model discussed in the preceding paragraph demonstrates that this band will give a reliable mass estimate: for example, if the recent star formation in the model is eliminated so there has been none for the last 100 Myr, the mass required to produce the same flux at 3.6 $\mu$m is only a factor of 1.6 higher. 

To evaluate the range of host galaxy properties and probable mass, we draw on  masses from modeling of galaxies with redshifts between 5 and 8 \citep{Labbe2010,Mclure2011,Curtis2013}. We adjust the IRAC Band 1 measurements from these references to z = 6.25 and K-correct them. The latter corrections are based on a model without the current starburst but allowing various levels of extinction, which we set according to the E(B-V) determined from the modeling in the cited papers. We impose a slope of 1, i.e. the mass should be proportional to the 3.6 $\mu$m luminosity. The data suggest a non-unity slope, but $\chi^2$ is only increased by a factor of 1.6 by forcing the slope to unity. The result is:

\begin{equation} \label{eq:massflux}
\log(M_*) = \log (F(\nu)) +10.05
\end{equation}

\noindent where the flux is in $\mu$Jy and $M_*$ in M$_\odot$. The scatter around this relation is 0.23 dex. The sample used for this calibration also gives insight to the expected level of extinction; 1 galaxy (3\%) has an indicated $A_V$ = 1.1, three more (10\%) have values consistent with $A_V =$ 1, and the rest (87\%) have indicated values of $A_V \le$ 0.5.

\begin{figure*}
    \centering
    \includegraphics[width=18cm]{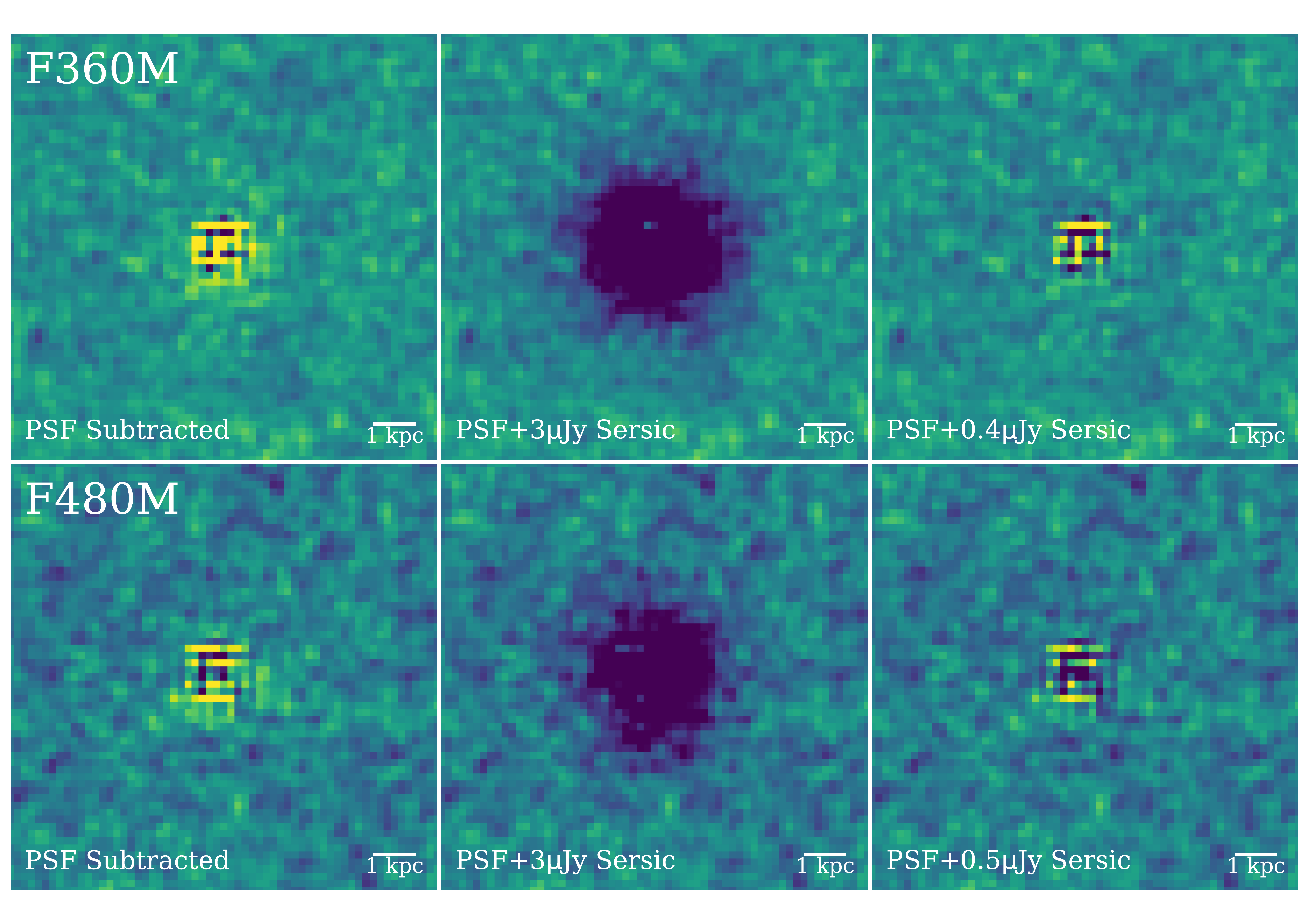}
    \caption{{\em left:} The PSF-subtracted F360M ({\em top}) and F480M ({\em bottom}) galaxy images, with the galaxy emission visible around the central PSF subtraction artifacts. When a S\'ersic profile with a measured flux of 3 $\mu$Jy---expected for a stellar mass of $10^{11}$ M$_{\odot}$---and effective radius of $\sim 1$ kpc is subtracted from each PSF-subtracted image {\em (center)}, the galaxy emission is strongly oversubtracted. Conversely, subtracting a profile with a flux of 0.4 $\mu$Jy (F360M) or 0.5$\mu$Jy (F480M) effectively removes the extended galaxy emission without evidence of oversubtraction {\em (right)}. This indicates that the stellar mass of the galaxy is likely closer to $10^{10}$ M$_{\odot}$.}
    \label{fig:sersic}
\end{figure*}

Equation \ref{eq:massflux} returns a galaxy mass of $\log{M_* \, (M_{\odot})} = 9.6 \pm 0.3$ from the F360M flux measured for the J2239+0207 host galaxy: this uncertainty includes both the uncertainty on the F360M flux from our sampling with a random set of apertures and the scatter in our empirical IRAC Band-1-stellar-mass relation. 
There are, however, a number of adjustments needed to this value. As already discussed, for consistency with our F210M upper limit, $A_V \ge 0.5$ is needed; we adopt $A_V$ = 1 at the rest wavelength of the F360M band (0.5 $\mu$m). This would put the host galaxy in the top 13\% of extinction among our calibrator sample. It would make the mass estimate $\log{M_* \, (M_{\odot})} = 10$. However, our model indicates that about 15\% of the flux in F360M comes from the H$\beta$ and [OIII] emission lines, which will be absent in the calibration galaxies since they have much lower levels of star formation. In addition, as discussed above, the contribution of the starburst in the host galaxy raises its flux in the F360M band by as much as a factor of 1.6 over that from less active galaxies such as those used for the calibration. Both of these effects would result in an overestimate of the host galaxy mass; we reflect this possibility with asymmetric error bars and quote the mass as $\log{M_* \, (M_{\odot})} = 10.0^{+0.3}_{-0.5}$.

The stellar mass is therefore significantly lower than the [C~II] dynamical mass estimated for the same galaxy by \cite{Izumi2019}, $\log{M_{\mathrm{dyn}} \, (M_{\odot})} \sim 11.5$. Oversubtraction of the PSF could suppress the measured flux and therefore the measured mass. However, using Equation \ref{eq:massflux} and adopting $A_V = 1$, as above, predicts a flux of 3 $\mu$Jy in F360M for $\log{M_*} = 11$ (increasing for less extinction): we can test this prediction by subtracting a simulated galaxy with this flux from our PSF-subtracted images. We create mock circular S\'ersic profiles (S\'ersic index $n=1$) with fluxes of 3 $\mu$Jy within a $0.4$\arcsec-radius aperture---identical to that used to extract the galaxy flux---and a range of effective radii, and subtract this mock galaxy profile from the PSF-subtracted quasar image. We find that regardless of the input effective radius, a 3 $\mu$Jy total flux invariably severely oversubtracts the galaxy emission (see Figure \ref{fig:sersic}, center panels). Conversely, a Se\'rsic profile with a $\sim 1$ kpc effective radius and total flux of $\sim0.4 \; \mu$Jy in F360M and $\sim0.5 \; \mu$Jy in F480M, when subtracted from the galaxy images, effectively removes any evidence of the host galaxy emission (Figure \ref{fig:sersic}, right panels). This Se\'rsic subtraction predicts total galaxy fluxes $\sim15$\% higher than those measured in the apertures in F360M and F480M, accounting for the fraction of the galaxy flux directly under the core of the PSF. 

Moreover, reducing the factor by which we scale the stellar PSF---therefore leaving more leftover flux after PSF subtraction and pushing the mass higher---cannot increase the mass enough to be consistent with the dynamical mass. Above a deviation of $\sim5\%$ from the best-fit PSF scale factor (at which point the resulting mass is still consistent within the errors with our original measurement) leftover PSF artifacts become obvious in the final PSF-subtracted images. A stellar mass of $\sim 10^{10}$ rather than $10^{11}$ $M_{\odot}$ is therefore reasonable.

Like many other quasars studied at similar redshifts \citep[e.g.][]{Izumi2021,Pensabene2020, Harikane2023}, J2239+0207 lies above the local $M_{\mathrm{BH}} - M_{\mathrm{bulge}}$ relation measured in elliptical galaxies and classical bulges by \cite{Kormendy2013} (see Figure \ref{fig:magorrian}). This picture may be influenced, however, by a number of both selection and measurement biases.

\subsection{Potential Biases in Host Galaxy Mass Measurements}

\subsubsection{Malmquist Bias in Quasar Sample}

When searching for high-redshift quasars via wide-field galaxy surveys, the most luminous quasars---with the most active black holes---are more readily observed; in addition, due to the rapid decline of the galaxy luminosity function at high masses, these active, massive black holes are more often found in more common, lower-luminosity, lower-mass galaxies (where they are outliers) rather than rarer, bright, higher-mass galaxies \citep[as discussed in][]{Lauer2007}. It has therefore been difficult to say with any certainty whether the observed tendency of high-redshift quasars to fall above the local $M_{\mathrm{BH}}-M_*$ relation, as J2239+0207 does, is evidence for evolution of $M_{\mathrm{BH}}/M_*$ with cosmic time, or simply a consequence of selection bias.

To evaluate this bias, we compare our derived mass for J2239+0207 and other measurements of $M_{\mathrm{BH}}/M_*$ at $z \sim 6$ to a local sample with the same biases. For 30 extremely luminous Palomar Green (PG) quasars (with $L_{\mathrm{AGN}} \geq 2 \times 10^{12} \, L_{\odot}$, at $z\leq0.5$), we retrieve black hole masses obtained via single-epoch estimations (i.e. using line widths and luminosities) from the literature \citep{Vestergaard2006, Veilleux2009, Yu2020}. In addition, we adopt total galaxy masses estimated from the quasar-galaxy image decomposition \citep{Zhang2016}. These considerations ensure that the black hole and galaxy masses for both this local sample and J2239+0207 are measured by similar methods. We plot these PG quasars alongside the $z \sim 6$ samples and J2239+0207 in Figure \ref{fig:magorrian}. By using the most luminous PG sources, we recreate the bias towards high $L_{\mathrm{AGN}}$ present in high-redshift quasar surveys; this sample choice also ensures that both samples are subject to the Malmquist-type bias in $M_{\mathrm{BH}}$ measurements derived from single-epoch spectra discussed by \cite{Shen2010}.

Despite both samples being similarly affected by the \cite{Lauer2007} and \cite{Shen2010} biases, the high-$L_{\mathrm{AGN}}$ PG quasars and $z \sim 6$ quasars still appear to follow different distributions in Figure \ref{fig:magorrian}. On average, the local quasars lie in more massive host galaxies at a given $M_{\mathrm{BH}}$ (the PG quasars are also largely consistent with the \cite{Kormendy2013} local $M_{\mathrm{BH}}-M_{\mathrm{bulge}}$ relation). The $z \sim 6$ quasars show much larger scatter, overlapping with the local examples but extending to a factor of $\sim$ 10 above them. 

\subsubsection{Use of Dynamical Masses of Host Galaxies}

J2239+0207 presents an example of a severely mismatched gas dynamical mass and stellar mass in a high-$z$ quasar. As JWST observations create larger samples of $z \gtrsim 6$ quasar host galaxy stellar masses, is a trend likely to emerge? Do we expect [C~II] gas dynamical masses to lie systematically above masses measured from stellar properties? 

 This appears not to be an issue up to moderate redshifts. In these cases, the dynamical masses come primarily from stellar spectroscopy, with some from MASER observations, and agree very well with stellar masses as measured in the $K_S$ band \citep{Kormendy2013}. Using either indicator for stellar mass, there is a tight relation between galaxy bulge mass and SMBH mass among non-AGN galaxies, often termed the `Magorrian Relation' \citep{Kormendy2013}. Many quasars have reliable stellar masses based on photometric and dynamical data; for $z \le 2$, their host galaxies also follow the local Magorrian Relation \citep{Kormendy2013}. 

\begin{figure}
    \centering
    \includegraphics[width=8.5cm]{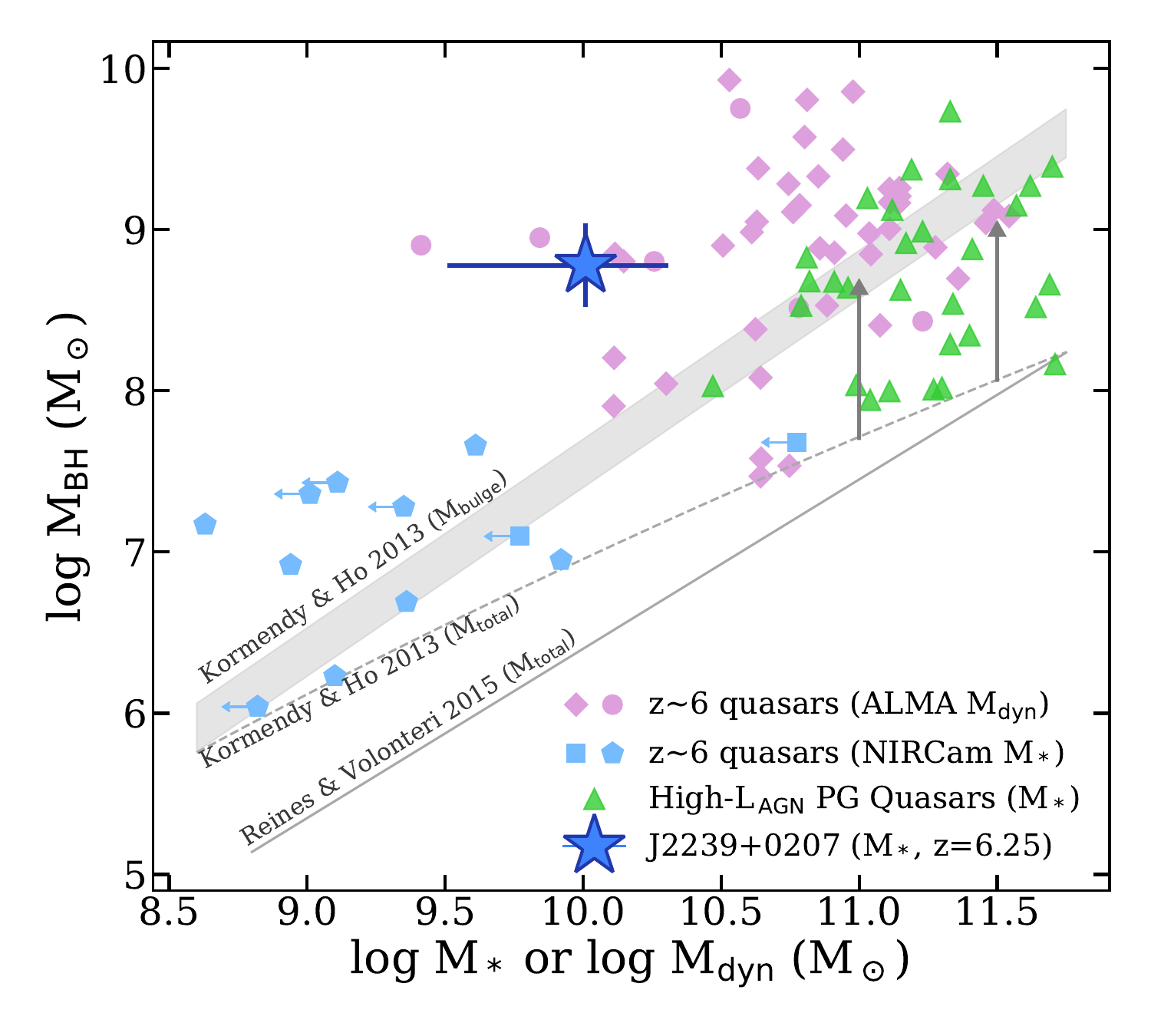}
    \caption{The $M_{\mathrm{BH}} - M_*$ relation for J2239+0207 (blue star) and a selection of other $z \sim 6$ and local quasars from the literature. $z \sim 6$ quasar dynamical masses are shown in grey (\cite{Izumi2021}, diamonds; \cite{Pensabene2020}, circles), while high-$z$ quasar stellar masses from recent JWST samples (\cite{Kocevski2023}, squares; \cite{Harikane2023}, pentagons) are shown in blue. Luminous Palomar Green (PG) quasars at $z<0.5$ with $L_{AGN} > 10^{12} L_{\odot}$ are shown as green triangles. The local $M_{\mathrm{BH}} - M_{\mathrm{bulge}}$ relation in local non-AGN systems from \cite{Kormendy2013} (grey shaded region) and the corresponding relation for the {\em total} stellar  mass (grey dashed line, masses scaled in proportion to $M_K$) are shown. Two vertical arrows indicate the Malmquist bias we find for the luminous PG quasars. We also show the $M_{\mathrm{BH}} - M_*$ relation for low-mass AGN host galaxies from \citet{Reines2015} as a grey solid line. PG quasars and $z \sim 6$ quasars appear to follow different distributions, with low-z quasars containing more massive host galaxies for a given black hole mass; however, a number of biases may affect this conclusion.}
    \label{fig:magorrian}
\end{figure}

However, this is a general issue at high redshift. Studies of the pre-JWST high-$z$ quasar samples \citep[including][gray points in Figure \ref{fig:magorrian}]{Izumi2021, Pensabene2020} compare the galaxy dynamical mass determined mostly from the [C~II] 158 $\mu$m line measured with ALMA with the local relation between galaxy bulge and black hole masses. As shown in Figure~\ref{fig:magorrian} and many other works,  at high redshift there is a large scatter relative to the local relation, with many galaxies indicated to have SMBH masses substantially in excess of its prediction. Is this effect an indication of a real change in the behavior of galaxies or is it a product of using the [C~II]-based dynamical masses rather than the approaches used at lower redshift? 

If the inferred dynamical masses tend to strongly {\em underestimate} stellar masses, the observed differences between local and high-$z$ quasars in Figure \ref{fig:magorrian} may simply stem from the resulting bias. On the other hand, if the dynamical masses {\em overestimate} the stellar mass, the trend in Figure \ref{fig:magorrian} is reinforced. 
The usual formalism \citep[e.g.,][]{Izumi2019} to translate line widths to dynamical masses is summarized in equation~\ref{eqn3}:

\begin{equation}
\frac{M_{\mathrm{dyn}}}{M_\odot} = 1.16 \times 10^5 \left( \frac{v_{\mathrm{circ}}}{\mathrm{km ~s}^{-1}}\right)^2 ~ \left( \frac{D}{\mathrm{kpc}} \right)
\label{eqn3}
\end{equation}

\noindent
where $D$ is the diameter of the galaxy and 

\begin{equation}
   v_{\mathrm{circ}} = 0.75 \frac{\mathrm{FWHM}}{ \sin i} 
\end{equation}

\noindent
with $i$ the angle of inclination, which is typically deduced from the ratio of the major and minor axes. The potential for {\it overestimation} lies primarily in the assumption that the line FWHM is dominated by the rotation of a disk galaxy. It is possible that non-Keplerian motions, e.g. winds, may be further broadening the line profile and biasing the dynamical mass higher \citep[see][]{Izumi2019}. Another possibility is that spheroidal-dominated galaxies with axial ratios close to unity will instead be assigned small inclinations appropriate for face-on disks, leading to a large and incorrect upward adjustment in $v_{\mathrm{circ}}$. The potential for {\it underestimation} lies in assigning too small an inclination, $i$, or diameter, $D$. A detailed theoretical simulation carried out by \cite{Lupi2019} found that the [C~II] dynamical mass {\it underestimates} the total stellar mass of a simulated rotationally supported galaxy at $z \sim 7$ by a factor of $2 - 5$, and potentially by more for a dispersion-dominated case. As they conclude, corrections of this size in the [C~II]-derived masses would substantially reduce the difference between local and high redshift systems shown in Figure \ref{fig:magorrian}. 

However, the dynamical mass for J2239+0207 from [CII] observations is $2.9 \times 10^{11}$ M$_\odot$ \citep{Izumi2019}; contrary to the models of \citet{Lupi2019}, this is nearly 30 times {\it larger} than our inferred stellar mass, despite the latter being determined in the rest-optical where the effects of extinction and recent star formation are minimized. At least in this case, it seems that the effects that can increase the estimated dynamical mass -- winds, incorrect values for inclination and diameter -- may be dominating the [C~II] line width  interpretation. 

We will evaluate this issue as follows. We will compare the dynamical masses based  on submm- and mm-wave line measures with those from photometry for the high $L_{AGN}$ PG quasars that are analogous to the high redshift quasars. This should show if the two methods yield consistent results, as is an implicit assumption in the comparison of the Magorrian ratios at high redshift with the local relation.

 We use CO line measurements of the PG quasar sample in place of [C~II]. The velocity widths of the [C~II] line and the CO lines are very similar in local galaxies \citep{rollig2012,deblok2016}, so we take the CO measurements as a proxy for [C~II] in the PG quasars. All the available CO measurements of PG quasars that also have direct stellar mass determinations \citep{Zhang2016} are listed in Table~\ref{tab:CO}. 
 We use equation (3) to translate the line widths to masses, taking the diameter of the host from \citet{Guyon2006,Veilleux2009}, with the former used for total diameters and the latter for bulges (the distinction between total and bulge is undetermined for high redshift host galaxies). The axial ratios for the PG quasars are generally close to 1 - the average for all but two where they are not available is 0.76. This would imply small inclinations and large corrections to the mass, but to be conservative we have set $\sin i$ = 1, and consequently our mass estimates are lower limits. Since our concern is that the masses might be too low, this does not undermine our analysis. 

As shown in Table~\ref{tab:CO}, assuming bulge diameters there are 3/10 cases where the dynamical mass lower limit is more than two times less than the directly determined stellar mass, and 3/10 cases where it is more than two times greater. When the dynamical masses are calculated using the total diameters of the galaxies, all ten cases have dynamical mass lower limits greater than the directly determined stellar masses. That is, the behavior of the PG quasars does not support the hypothesis that the line-widths-approach results in a significant and uniform underestimate of host galaxy masses. 

This analysis may point to a genuine evolution in $M_{\mathrm{BH}}/M_*$ with cosmic time, rather than an apparent evolution solely due to biases. However, the scatter in Figure~\ref{fig:magorrian} suggests that any such evolution will not appear uniformly. Instead, the $M_{\mathrm{BH}}$-$M_*$ ratio is likely to take on a much larger range at high redshift than occurs locally. Many more quasar host galaxies will be directly measured with JWST in the coming years, and will help probe this possibility.  

\begin{deluxetable*}{lccccccccc}
\tabletypesize{\footnotesize}
\tablecaption{Comparison of Mass Determinations}
\tablewidth{0pt}
\tablehead{
\colhead {ID} &
\colhead {CO FWHM} &
\colhead {ref\tablenotemark{a}}  &
\colhead{log(M$_*$/M$_\odot$)\tablenotemark{b}}  &
\colhead {r$_{1/2, \mathrm{bulge}}$\tablenotemark{c}}  &
\colhead{log(M$_{\mathrm{dyn}}$/M$_\odot$)\tablenotemark{d}}  &
\colhead {M$_{\mathrm{dyn}}$/M$_*$\tablenotemark{e}} &
\colhead {r$_{\mathrm{total}}$\tablenotemark{f}}  &
\colhead{log(M$_{\mathrm{dyn}}$/M$_\odot$)\tablenotemark{g}}  & 
\colhead{M$_{\mathrm{dyn}}$/M$_*$\tablenotemark{h}}  \\
\colhead {} &
\colhead {(km s$^{-1}$)} &
\colhead {}  &
\colhead{}  &
\colhead {(kpc)}  &
\colhead{(lower limit)}  &
\colhead {(lower limit)} &
\colhead {(kpc)}  &
\colhead{(lower limit)}  & 
\colhead{(lower limit)}  
}
\startdata
PG 0007+106	&	387	&	1	&	11.02	&	2.97	&	11.76	&	5.54	&		&		&		\\
PG 0050+124	&	378	&	1	&	11.3		&	1.03	&	11.28	&	0.96	&		&		&		\\
PG 0838+770	&	90	&	2	&	11.32	&	0.56	&	9.77	&	0.03	&		&		&		\\
PG 0923+129	&	362	&	1	&	11.27	&		&		&		&		&		&		\\
PG 1119+120	&	213	&	1	&	10.85	&	0.45	&	10.42	&	0.38	&	1.47	&	10.94	&	1.23	\\
PG 1126-041	&	467	&	1	&	11.03	&	0.79	&	11.35	&	2.10	&	3.08	&	11.94	&	8.19	\\
PG 1211+143	&	66	&	1	&	10.56	&		&		&		&	11.50	&	10.81	&	1.79	\\
PG 1229+204	&	202	&	1	&	11.12	&	3.65	&	11.29	&	1.48	&	4.16	&	11.35	&	1.68	\\
PG 1426+015	&	344	&	1	&	11.23	&	1.64	&	11.40	&	1.49	&	8.70	&	12.13	&	7.90	\\
PG 1440+356	&	370	&	2	&	11.23	&	0.39	&	10.84	&	0.41	&	1.53	&	11.44	&	1.61	\\
PG 1613+658	&	490	&	2	&	11.64	&		&		&		&	15.23	&	12.68	&	10.93	\\
PG 1700+518	&	260	&	3	&	11.57	&		&		&		&	18.78	&	12.22	&	4.46	\\
PG 2130+099	&	548	&	1	&	11.03	&	2.83	&	12.05	&	10.36	&	1.53	&	11.78	&	5.61	\\
PG 2214+139	&	180	&	1	&	11.16	&	2.78	&	11.07	&	0.81	&	3.45	&	11.16	&	1.01	\\
\enddata
\tablenotetext{a}{1 -- \citet{shangguan2020}: 2 -- \citet{Evans2001}: 3 -- \citet{Evans2009}}
\tablenotetext{b}{From stellar photometry, \citet{Zhang2016}.}
\tablenotetext{c}{From \citet{Veilleux2009}, Table 5}
\tablenotetext{d}{Based on computing the dynamical mass for r$_{\mathrm{bulge}}$. These values are lower limits because sin$i$ has been set to 1. }
\tablenotetext{e}{Lower limits, based on masses for r$_{\mathrm{bulge}}$.}
\tablenotetext{f} {The larger of the bulge and disk components fitted by \citet{Guyon2006}, table 6.}
\tablenotetext{g}{Lower limit because sin$i$ is set to 1, based on dynamical mass computed for r$_{\mathrm{total}}$.}
\tablenotetext{h}{Lower limits, based on masses for r$_{\mathrm{total}}$.}
\label{tab:CO}
\end{deluxetable*}

\section{Summary and Conclusions} \label{sec:summary}

We utilize dedicated observations of a PSF star in three JWST/NIRCam medium bands, as well as a NIRSpec prism spectrum, to remove the contribution from the central point source of the quasar J2239+0207 and characterize its host galaxy. 

\begin{itemize}
\itemsep -0.25em
    \item Using the broad H$\alpha$ line in the NIRSpec spectrum, we obtain a black hole mass of $3.4 \times 10^8$ M$_{\odot}$, somewhat  lower than previously estimated values from Mg II and [C IV]. 
    \item We subtract the PSF in our three bands and detect emission from the host galaxy in the two filters longward of the Balmer break, measuring host galaxy fluxes of $0.34 \; \mu$Jy in F360M and $0.44 \; \mu$Jy in F480M (host-to-quasar flux ratios of 4.6\% and 3.9\% respectively). We relate the F360M flux to galaxy mass using IRAC Band 1 observations of galaxies from $5<z<8$ in the literature, obtaining a stellar mass of approximately $10^{10}$ M$_{\odot}$, which is significantly less than the existing dynamical mass measurement from 158 $\mu$m [C~II]. J2239+0207 lies above the local $M_{\mathrm{BH}}-M_{*}$ relation, and is among the lower-mass host galaxies explored at $z \sim 6$.
    \item To explore the $M_{\mathrm{BH}}-M_{*}$ relation of J2239+0207 and other $z\sim6$ quasars and its deviation from the local Universe, we utilize a sample of high-luminosity PG quasars with masses from single-epoch spectra and total host galaxy masses derived from photometry. This selection ensures that both of our samples are subject to the \cite{Lauer2007} bias, and any evolution with redshift is not a result of this bias. The host of J2239+0207 is significantly undermassive compared with typical local quasar hosts. 
    \item At a given $M_{\mathrm{BH}}$, the $z\sim6$ quasars from the literature appear to show a much larger range of host galaxy masses than found in local cases. Most of the $z\sim6$ quasars use [C~II] dynamical mass as a proxy for stellar mass. We evaluate the determination of host galaxy masses through this method by comparing with similar measurements using CO lines from PG quasars. We find no evidence for systematic underestimates that might result in the relatively low host masses deduced for many high redshift systems. We hypothesize that the $M_{\mathrm{BH}}$-$M_*$ ratio will possess more scatter at higher redshift than in the local Universe, but larger samples are required to confirm this behavior.
\end{itemize}

JWST's wavelength range and sensitivity will for the first time allow the detection and characterization of large numbers of high-redshift quasar hosts in the rest-frame optical, using NIRCam photometry. The method of PSF subtraction using observations of a dedicated PSF star is able to retrieve galaxy signals even at the $<5$\% host-to-quasar level. As stellar masses are obtained and compared to [C~II] dynamical masses for more high-redshift quasars and local and high-redshift samples can be compared consistently, our analysis of J2239+0207 demonstrates that JWST will allow for the robust characterization of any evolution of the $M_{\mathrm{BH}}-M_{*}$ relation over cosmic time. 

\begin{acknowledgements}
We thank Tod Lauer for important discussions regarding selection biases and Feige Wang, Jinyi Yang, and Xiaohui Fan for their insightful comments. The anonymous referee's feedback also greatly improved this work. MS, SA, JL, and GR acknowledge support from the JWST Mid-Infrared Instrument (MIRI) grant 80NSSC18K0555, and the NIRCam science support contract NAS5-02105, both from NASA Goddard Space Flight Center to the University of Arizona. The JWST data presented in this paper were obtained from the Mikulski Archive for Space Telescopes (MAST) at the Space Telescope Science Institute. The specific observations analyzed can be accessed via \dataset[DOI.]{http://dx.doi.org/10.17909/ajqw-7w52}
\end{acknowledgements}

\vspace{5mm}

\facilities{JWST(NIRCam, NIRSpec)}

\software{Astropy \citep{Astropy2013,Astropy2018},  
         Matplotlib \citep{Hunter2007}, NumPy \citep{VanDerWalt2011}, photutils \citep{Bradley2022}, 
         WebbPSF \citep{Perrin2014}
         }

\bibliography{bibliography}{}
\bibliographystyle{aasjournal}

\end{document}